# Photothermally Excited Contact Resonance Imaging in Air and Water

Marta Kocun, Aleksander Labuda, Anil Gannepalli and Roger Proksch.
*Asylum Research, an Oxford Instruments Company, Santa Barbara, CA, USA*

Contact Resonance Force Microscopy (CR-FM) is a leading AFM technique for measuring viscoelastic nano-mechanical properties. Conventional piezo-excited CR-FM measurements have been limited to imaging in air, since the "forest of peaks" frequency response associated with acoustic excitation methods effectively masks the true cantilever resonance. Using photothermal actuation results in clean contact resonance spectra, that closely match the ideal frequency response of the cantilever, allowing unambiguous and simple resonance frequency and quality factor measurements in air and liquids alike. This extends the capabilities of CR-FM to biologically relevant and other soft samples in liquid environments. We demonstrate CR-FM in air and water on both stiff silicon/titanium samples and softer polystyrene-polyethylene-polypropylene polymer samples with the quantitative moduli having very good agreement between expected and measured in both environments.

The use of atomic force microscopy (AFM) as a nanoscale material characterization tool has increased in the last two decades due to its versatility in generating contrast from a variety of physical properties, such as electrical, magnetic, morphological and mechanical properties. While AFM routinely provides contrast, acquiring quantitative measurements is notoriously difficult.[1–7]

This paper focuses on contact resonance CR-FM, a technique based on dynamic contact mode of AFM, and originally developed to measure elastic properties of stiff materials.[8–11] In the CR-FM techniques, vibrational resonances of the AFM cantilever are excited while the tip is in contact with the sample. A notable advantage of CR techniques over static deflection based techniques is a higher sensitivity to contact stiffness as it measures frequency shifts of the cantilever and has better signal-to-noise from resonance amplification. CR techniques are implemented by mechanical actuation of either the sample (atomic force acoustic microscopy, AFAM)[10] or the cantilever base (ultrasonic AFM, UAFM).[11] Recently, CR-FM has been extended to measure viscoelastic properties of materials.[12–14] In this technique, henceforth referred to as viscoelastic CR-FM, the contact resonance frequency is tracked during scanning and the changes in frequency and quality factor are used to generate a map of storage and loss moduli using theoretical models.

To date, CR-FM has rarely been employed in aqueous media, in part due to difficulties associated with obtaining clean contact cantilever transfer functions.[15,16] Furthermore, characterization of CR-FM results obtained in liquid environments requires additional analysis.[17–19] Measuring mechanical properties in liquid offers the advantage of eliminating capillary adhesion forces present during air imaging. Additionally, many samples, most notably biological ones, require hydration to preserve their native state for mechanical property measurements.[20,21]

Here, we demonstrate photothermally actuated[22] viscoelastic CR-FM in air and water over a wide range of moduli.[23] The samples investigated included very stiff materials (~100-200 GPa) – a silicon wafer with a titanium metal evaporated film, and a softer (~2-3 GPa) ternary polymer blend composed of polypropylene, polyethylene, and polystyrene. The results are compared with measurements obtained using conventional piezo-acoustic sample actuation results.

All Contact Resonance Force Microscopy (CR-FM) measurements were conducted on a Cypher AFM (Asylum Research, Santa Barbara, CA) equipped with a contact resonance sample actuator and a blueDrive™ photothermal actuation module. While imaging in liquid, a droplet cantilever holder was used with 18.2 MΩ water. All cantilevers had a nominal spring constant of 2 N/m (AC240 Cantilevers, Olympus Corporation, Tokyo, Japan) with a free cantilever resonance of approximately 70 kHz and a corresponding first flexural mode contact resonance occurring in a range of 260-300 kHz. To allow for a direct comparison between the two drive methods, a contact tune was first acquired using the sample actuator, immediately followed by contact tune using photothermal actuation. The resonance frequency was tracked using dual AC resonance tracking (DART) mode, a technique where the cantilever is maintained at its contact resonance frequency by monitoring the amplitude and phase at two frequencies bracketing the contact resonance.[24,25]

To quantify viscoelastic properties, the tip-sample contact is modeled as a driven damped harmonic oscillator, which enables us to calculate the resonance frequency $f$ and quality factor $Q$.[11–13,26] The maps of these parameters are then used to obtain viscoelastic material properties as outlined below and discussed in more detail elsewhere.[13,14,26]

A schematic representation of the photothermal cantilever actuation setup is shown in Figure 1a. As the sample is scanned in contact mode, the infrared laser is used to measure the cantilever deflection (including the dynamic response). The low amplitude cantilever oscillation is induced by modulating the blue laser power that is focused at the base of the cantilever.

Representative contact resonance frequency spectra of a silicon probe on a silicon surface are shown in Figure 1 b-c, where the piezoacoustic actuation and the photothermal actuation responses are plotted. The results in air for both actuation methods are very similar. The advantage of using photothermal actuation becomes clear when the contact resonance tunes are performed in water, as shown in Figure

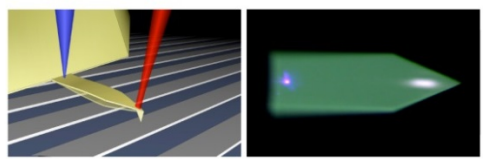

a) Diagram and camera view of a photothermally driven cantilever

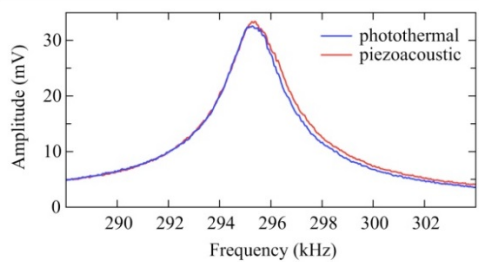

b) Contact tune in air

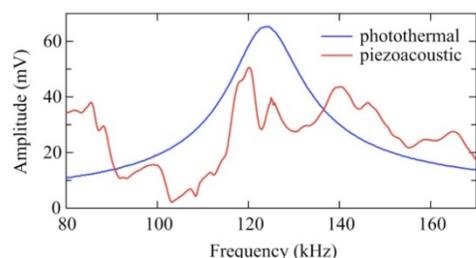

c) Contact tune in water

FIG. 1. (a) Schematic representation of photothermal cantilever actuation during contact resonance imaging. In air (b), contact resonance frequency for piezoacoustic actuation was 295.3 kHz, Q = 138. For photothermal actuation the contact resonance frequency was 295.4 kHz, Q = 148. In water (c), piezoacoustically actuated spectrum was complex with multiple peaks while the photothermally driven tune showed a single clean peak at 124.1 kHz, Q = 8.1.

1(c). In contrast to the clean single contact resonance peak obtained by photothermal actuation, the shape of the contact resonance peak obtained with piezoacoustic actuation is plagued by the "forest of peaks".[15,27–29]

Figure 2 shows a 5μm × 10μm area of a titanium film evaporated on silicon sample that was first imaged in air and subsequently in water; both images were acquired using photothermal excitation.

Topography images (figures 2a and 2a') reveal ~200 nm high, rough titanium stripes on a smooth silicon surface. The contact resonance frequency (figure 2b) is higher (~290 kHz) on silicon surface in comparison to titanium (~280 kHz), which correlates well with the expected higher silicon stiffness. The frequency contrast remains the same (higher on silicon, lower on titanium) regardless of the imaging conditions (air or water). Images of $Q$ factor values, associated with energy dissipation, are shown in figures 2c and 2c'. $Q$ values observed on titanium were higher than those measured on silicon. The expectation for the energy dissipation to be lower on silicon was not observed in our experiments. Squeeze-film damping due to the ~200 nm height difference between the silicon and titanium and other considerations beyond the scope of this work may effectively lower the $Q$ values observed on silicon.[19]

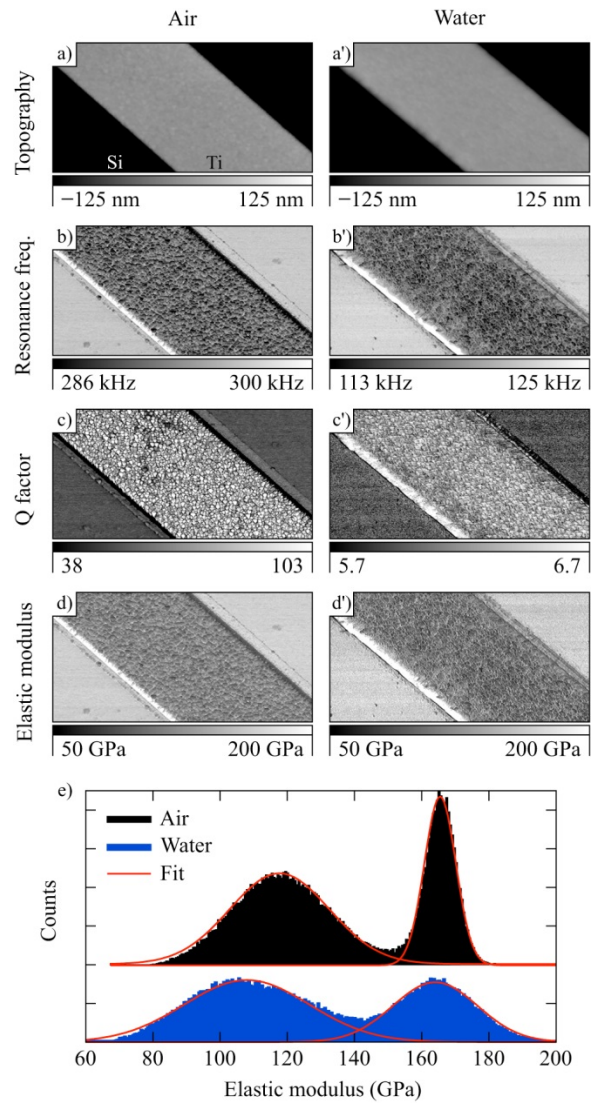

FIG. 2. Contact resonance images (5μm × 10μm) of a titanium film evaporated onto a silicon surface. (a) and (a') show the topography of the sample in air and water, respectively. (b) and (b') show contact resonance frequency of the cantilever in air and water, respectively. (c) and (c') show $Q$ factor images in air and water, respectively. (d) and (d') show the calculated elastic modulus ($E'$) images in air and water respectively. In (e), histograms of the $E'$ obtained in air (top) and water (bottom). The loading forces were ~800 nN in air and ~700 nN in water.

Calculated storage moduli ($E'$) of silicon and titanium, in air and water, are shown in figure 2d and 2d'. The results presented here were obtained using internal calibration approach where silicon was chosen as the reference material and its median $E'$ value was assumed to correspond to 165 GPa. The expected and calculated $E'$ values for all the materials studied are summarized later in the text, in table 1.

Next, we measured a soft "ternary" blended polymer composed of polypropylene (PP), polyethylene (PE), and polystyrene (PS) with expected moduli ranging between 2 and 3 GPa (see Table 1)[30] using photothermal CR-FM in air and water. Figure 3a shows the air topography of the sample. The topography image acquired in water (figure 3a') shows improved imaging resolution presumably due to reduced tip-sample adhesion.

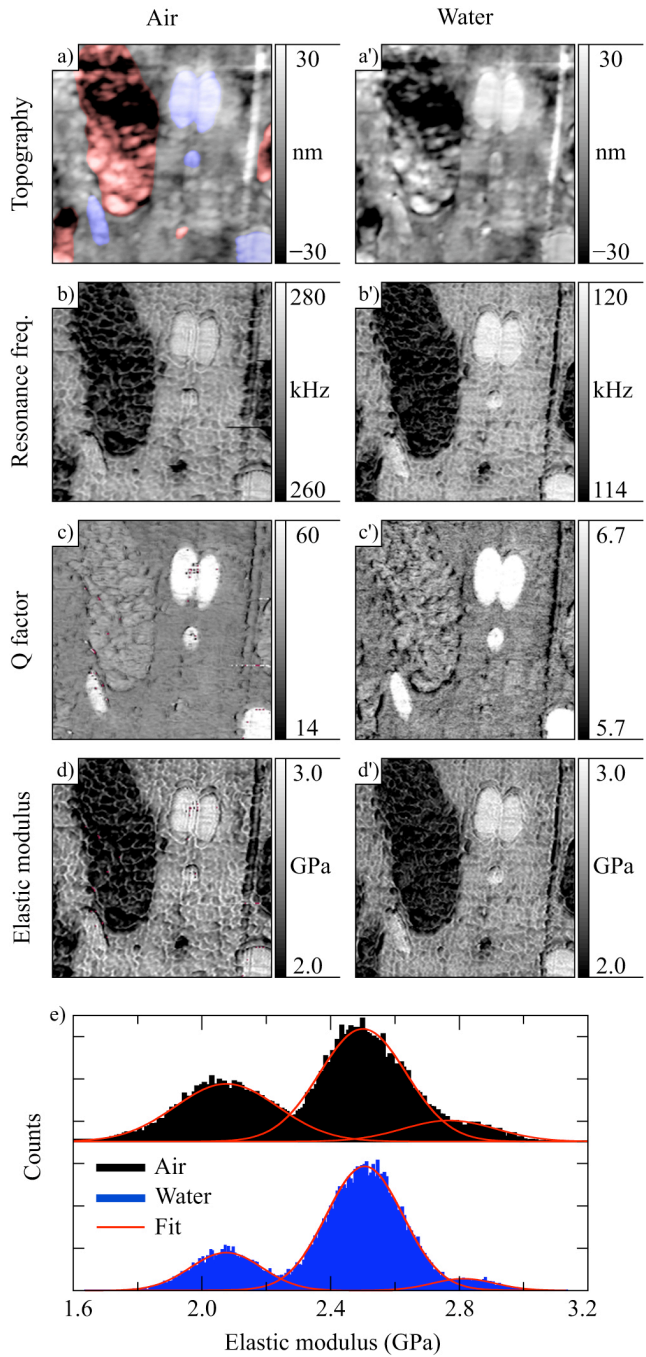

FIG 3: Contact resonance images (5μm × 10μm) of a ternary polymer blend composed of polyethylene (PE – red tint), polypropylene (PP – matrix) and polystyrene (PS – blue tint). (a) and (a') show the topography of the sample in air and water, respectively. (b) and (b') show contact resonance frequency of the cantilever in air and water, respectively. (c) and (c') show $Q$ factor images in air and water, respectively. In (e), histograms of the $E'$ values obtained in air (top) and water (bottom). The loading forces were ~450 nN in air and ~130 nN in water.

The contact resonance frequency values shown in Figures 3b and b' were highest on PS and lowest values on PE, as expected. Figure 3c and 3c' show the $Q$ factor images. PS inclusions are well distinguished from PE and PP domains, however, the two latter components show very little contrast between each other, indicating their relatively similar dissipation. Figure 3e, compares histograms of $E'$ values calculated from the images performed in air and in water, clearly differentiating the three components.

As with the stiffer sample, an internal standard was used for the polymer, where the matrix material (PP) was set to be the reference standard with $E'$ of 2.5 GPa.[30] Images showing the calculated storage moduli and the associated histograms for each component are shown in Figures 3d, 3d' and 3e. Table 1 summarizes all of the calculated storage moduli for the materials investigated in this study, in air and water. The three components of the sample were clearly distinguished between each other and their storage moduli exhibited the same elasticity trend as reported previously where PE < PP < PS.

| Sample | $E'$ measured in air (GPa) | $E'$ measured in water (GPa) | Expected $E'$ values[*] (GPa) |
|---|---|---|---|
| Si | 165.5 ($\sigma$ =4.2) | 165.7 ($\sigma$ =13.3) | 165 |
| Ti | 121.5 ($\sigma$ =14.3) | 106.5 ($\sigma$ =17.1) | 110-125 |
| PE | 2.08 ($\sigma$ =0.02) | 2.08 ($\sigma$ =0.11) | 2.12 |
| PP | 2.50 ($\sigma$ =0.13) | 2.51 ($\sigma$ =0.12) | 2.50 |
| PS | 2.77 ($\sigma$ =0.14) | 2.81 ($\sigma$ =0.10) | 2.84 |

Table 1. Summary of elastic modulus ($E'$) values estimated in this work and expected values.[30,31] Si and PP are used as reference standards.
*Values are obtained from macroscopic bulk DMA values calculated at 250 kHz via time-temperature superposition.[30]

The agreement between the previously reported values[30] and the results shown here demonstrate the efficacy of photothermal actuation for CR imaging. Both stiff (silicon/titanium) and softer materials (polymer blend) were successfully imaged and analyzed providing storage moduli values using CR-FM with photothermal excitation. Strong contrast was observed in the frequency data when imaged in water, even for very similar materials. In the future, quantitative CR-FM maps of biological tissues may prove valuable for cells, proteins and biomaterial engineering of materials, such as implants, that are primarily used in liquid settings.

**Acknowledgments**

The authors gratefully acknowledge Donna C. Hurley (NIST) for providing the Ti/Si sample and Dalia Yablon and Andy Tsou at Exxon Mobil Corporate Strategic Research for providing the ternary blend sample.

**References**

[1] C.J. Gómez and R. Garcia, Ultramicroscopy **110**, 626 (2010).

[2] N.F. Martínez and R. García, Nanotechnology **17**, S167 (2006).

[3] M. Radmacher, R.W. Tillmann, and H.E. Gaub, Biophys. J. **64**, 735 (1993).

[4] A. Rosa-Zeiser, E. Weilandt, S. Hild, and O. Marti, Meas. Sci. Technol. **8**, 1333 (1997).

[5] M.E. Dokukin and I. Sokolov, Langmuir **28**, 16060 (2012).

[6] D. Platz, E.A. Tholén, D. Pesen, and D.B. Haviland, Appl. Phys. Lett. **92**, 153106 (2008).

[7] D. Wang, S. Fujinami, K. Nakajima, and T. Nishi, Macromolecules **43**, 3169 (2010).

[8] D.C. Hurley, *Contact Resonance Force Microscopy Techniques for Nanomechanical Measurements in Applied Scanning Probe Methods Vol XI* (Springer-Verlag, Berlin, 2009), pp. 97–138.

[9] D.C. Hurley, K. Shen, N.M. Jennett, and J.A. Turner, J. Appl. Phys. **94**, 2347 (2003).

[10] U. Rabe and W. Arnold, Appl. Phys. Lett. **64**, 1493 (1994).

[11] K. Yamanaka, H. Ogiso, and O. Kolosov, Appl. Phys. Lett. **64**, 178 (1994).

[12] A. Gannepalli, D.G. Yablon, A.H. Tsou, and R. Proksch, Nanotechnology **22**, 355705 (2011).

[13] J.P. Killgore, D.G. Yablon, A.H. Tsou, A. Gannepalli, P.A. Yuya, J.A. Turner, R. Proksch, and D.C. Hurley, Langmuir **27**, 13983 (2011).

[14] P.A. Yuya, D.C. Hurley, and J.A. Turner, J. Appl. Phys. **104**, 74916 (2008).

[15] T.E. Schäffer and P.K. Hansma, J. Appl. Phys. **84**, 4661 (1998).

[16] D. Kiracofe and A. Raman, Nanotechnology **22**, 485502 (2011).

[17] J.E. Sader, J. Appl. Phys. **84**, 64 (1998).

[18] N. Ploscariu and R. Szoszkiewicz, Appl. Phys. Lett. **103**, 263702 (2013).

[19] R.C. Tung, J.P. Killgore, and D.C. Hurley, Rev. Sci. Instrum. **84**, 73703 (2013).

[20] S.E. Campbell, V.L. Ferguson, and D.C. Hurley, Acta Biomater. **8**, 4389 (2012).

[21] S. Hengsberger, A. Kulik, and P. Zysset, Bone **30**, 178 (2002).

[22] A. Labuda, J. Cleveland, N. Geisse, M. Kocun, B. Ohler, R. Proksch, M. Viani, and D. Walters, Microsc. Anal. **28**, 23 (2014).

[23] A. Labuda, K. Kobayashi, Y. Miyahara, and P. Grütter, Rev. Sci. Instrum. **83**, 053703 (2012).

[24] B.J. Rodriguez, C. Callahan, S. V Kalinin, and R. Proksch, Nanotechnology **18**, 475504 (2007).

[25] US Patents: 8 024 963, 7 937 991, 7 603 891, 7 921 466, 7 958 563

[26] P.A. Yuya, D.C. Hurley, and J.A. Turner, J. Appl. Phys. **109**, 113528 (2011).

[27] A. Labuda, K. Kobayashi, D. Kiracofe, K. Suzuki, P.H. Grütter, and H. Yamada, AIP Adv. **1**, 022136 (2011).

[28] R. Proksch and S. V Kalinin, Nanotechnology **21**, 455705 (2010).

[29] T.E. Schäffer, J.P. Cleveland, F. Ohnesorge, D.A. Walters, and P.K. Hansma, J. Appl. Phys. **80**, 3622 (1996).

[30] D.G. Yablon, A. Gannepalli, R. Proksch, J. Killgore, D.C. Hurley, J. Grabowski, and A.H. Tsou, Macromolecules **45**, 4363 (2012).

[31] R. Boyer, G. Welsch, and E.W. Collings, editors, *Materials Properties Handbook Titanium Alloys* (ASM International, Materials Park, 1994), p. 131.